\documentstyle[emulateapj,psfig]{article}

\def\ale{\mathrel{\hbox{\rlap{\hbox{\lower4pt\hbox{$\sim$}}}\hbox{$<$}}}}
\def\age{\mathrel{\hbox{\rlap{\hbox{\lower4pt\hbox{$\sim$}}}\hbox{$>$}}}}

\def\arcmin{\hbox{$^\prime$}}

\def\gsim{\mathrel{\hbox{\rlap{\lower.55ex \hbox {$\sim$}}
                   \kern-.3em \raise.4ex \hbox{$>$}}}}
\def\lsim{\mathrel{\hbox{\rlap{\lower.55ex \hbox {$\sim$}}
                   \kern-.3em \raise.4ex \hbox{$<$}}}}

\def\aa#1#2#3{\bibitem[]{}#1, A\&A, #2, #3.}
\def\aasup#1#2#3{\bibitem[]{}#1, A\&AS, #2, #3.}

\def\apj#1#2#3{\bibitem[]{}#1, {\it Ap. J.}, {\bf#2}, #3.}
\def\apjlett#1#2#3{\bibitem[]{}#1, {\it Ap. J. (Letters)}, {\bf #2},
#3.}

\def\araa#1#2#3{\bibitem[]{}#1, ARA\&A, #2, #3.}

\def\iauc#1#2{\bibitem[]{}#1, IAU Circ.~No.~#2}
\def\mnras#1#2#3{\bibitem[]{}#1, {\it M.N.R.A.S.}, {\bf#2}, #3.}
\def\nature#1#2#3{\bibitem[]{}#1, {\it Nature}, {\bf #2}, #3.}

\def\pasp#1#2#3{\bibitem[]{}#1, PASP, #2, #3.}

\lefthead{Bloom et al.}
\righthead{Spectroscopy of GRB 970508}
\begin{document}

\submitted{Submitted to {\it ApJ Letters} on 30 July 1998}
 
\title{The Host Galaxy of GRB 970508\footnotemark}

\footnotetext{Based on the observations obtained at the W.~M.~Keck
Observatory which is operated by the California Association for
Research in Astronomy, a scientific partnership among California
Institute of Technology, the University of California and the National
Aeronautics and Space Administration.}
 
\author{J.~S.~Bloom\altaffilmark{2}, S.~G.~Djorgovski,
S.~R.~Kulkarni}

\affil{California Institute of Technology, Palomar Observatory,
105-24, Pasadena, CA 91125}

\author{and}
\author{D.~A.~Frail}

\affil{National Radio Astronomy Observatory, P.~O.~Box O, Socorro, NM
87801} \altaffiltext{2}{email \tt{jsb@astro.caltech.edu}}

\begin{abstract}
We present late-time imaging and spectroscopic observations of the
optical transient (OT) of gamma-ray burst (GRB) 970508.  Imaging
observations roughly 200 and 300 days after the burst provide
unambiguous evidence for the flattening of the light-curve.  The
spectroscopic observations reveal two persistent features which we
identify with [O II] 3727 \AA\ and [Ne III] 3869 \AA\ at a redshift of
$z = 0.835$ --- the same redshift as the absorption system seen when
the transient was bright.  The OT was coincident with the underlying
galaxy to better than 370 milliarcsec or a projected radial separation
of less than 2.7 kpc.  The luminosity of the [O II] line implies a
minimum star-formation rate of  $\age 1 ~M_\odot ~{\rm yr}^{-1}$.  In
our assumed cosmology, the implied restframe absolute magnitude is
$M_B = -18.55$, or $L_{B} = 0.12 L_*$.  This object, the likely host
of GRB 970508, can thus be characterized as an actively star-forming
dwarf galaxy.  The close spatial connection between this dwarf galaxy
and the GRB requires that at least some fraction of progenitors be not
ejected in even the weakest galactic potentials.

\end{abstract}

\section{Introduction}

After an initial brightening lasting $\sim$1.5 days, the optical
transient (OT) of gamma-ray burst (GRB) 970508 faded with a nearly
pure power-law slope by 5 magnitudes over $\sim$100 days (eg.~Galama
et al.~1998; Garcia et al.~1998; Sokolov et al.~1998).  Indications of
a flattening in the light-curve (Pedersen et al.~1997) were confirmed
independently by Bloom et al.~1998, Castro-Tirado et al.~1998, and
Sokolov et al.~1998.

The existence of an [O II] emission line at the absorption system
redshift (Metzger et al.~1997b) was taken as evidence for an
underlying, dim galaxy host.  After HST images revealed the
point-source nature of the light, Fruchter, Bergeron, \& Pian~1997,
Pian et al.~1998, and Natarajan et al.~1997 suggested that the source
responsible for the \hbox{[O II]} emission must be a very faint ($R
\age 25$ mag), compact ($\ale 1$ arcsec) dwarf galaxy at $z=0.835$
nearly coincident on the sky with the transient.  These predictions
are largely confirmed in the present study.

In this {\it Letter} we report on the results of deep imaging and
spectroscopy of the host galaxy of GRB 970508 obtained at the 10-m
Keck II telescope.

\section{Observations and Analysis}

Imaging and spectroscopic observations were obtained using the Low
Resolution Imaging Spectrograph (LRIS; Oke et al.~1995) on the 10-m
Keck II Telescope on Mauna Kea, Hawaii.  The log of the observations
is presented in Table 1.  All nights were photometric.  The imaging
data were reduced in the standard manner.

To follow the light-curve behavior of the OT+host over $\sim$300 days from
the time of the burst, we chose to tie the photometric zero-point to a
previous study (Sokolov et al.~1998) which predicted late-time
magnitudes based on early ($\ale 100$ days) power-law behavior in
several bandpasses.  This provides an internally consistent data set
for our purposes. Other studies of the light-curve include Galama et
al.~1998 and Pian et al.~1998.  V.~Sokolov (private communication)
provided magnitudes of eight ``tertiary'' field stars ($R$ = 18.7 to
23 mag) as reference since the four secondary comparison stars
(Sokolov et al.~1998) were saturated in all our images.  The
zero-points were determined through a least squares fit and have
conservative errors $\sigma_B = 0.05$ and $\sigma_R = 0.01$ mag.

For our spectroscopic observations we used a 300 grating which gives a
typical resolution of $\approx 15$ \AA, and a wavelength range from
approximately 3900 to 8900 \AA.  The spectroscopic standards G191B2B
(Massey et al.~1988) and HD 19445 (Oke \& Gunn 1983) were used to flux
calibrate the data of October and November, respectively.  Spectra
were obtained with the slit position angle at $51^\circ$, in order to
observe both the host galaxy of GRB 970519, and g1.  This angle was
always close to the parallactic angle, and the wavelength-dependent
slit losses are not important for the discussion below.  Internal
consistency implied by measurements of independent standards implies
an uncertainty of less than 20\% in the flux zero point calibration.
Exposures of arc lamps were used for the wavelength calibration, with
a resulting r.m.s.~uncertainty of about 0.3 \AA, and possible
systematic errors of the same order, due to the instrument flexure.

\section{Results}

Table 1 gives a summary of the derived magnitudes at the position of
the OT and, as a comparison, the predicted magnitude from a pure-power
law decay fit by Sokolov et al.~(1998).  The OT+host is brighter by
$\age 0.8$ magnitudes in both $B$- and $R$-band leading to the obvious
conclusion that transient has faded to reveal a constant source.  We
used a Levenberg-Marquardt $\chi$-square minimization method to fit a
power law flux (OT) plus constant flux (galaxy) to the $B$ and
$R$ light-curves using data compiled in Sokolov et al.~(1998):
\begin{equation}
f_{\rm total} = f_{0}t^{-\alpha} + f_{\rm gal},
\end{equation}
where $t$ is the time since the burst measured in days.  The
quantities $f_{0}$ and $f_{\rm gal}$ are the normalization of the flux
of the transient and the persistent flux of the underlying galaxy,
respectively.  We find $B_{gal} = 26.77 \pm 0.35$ mag,~$R_{gal} =
25.72 \pm 0.20$ mag, $B_{0} = 19.60 \pm 0.04$ mag, $R_{0} = 18.79 \pm
0.03$ mag, and $\alpha_B = -1.31 \pm 0.03$, $\alpha_R = -1.27 \pm
0.02$.  The data are statistically inconsistent with a pure power-law
decline.

\begin{figure*}[tb]
\centerline{\psfig{file=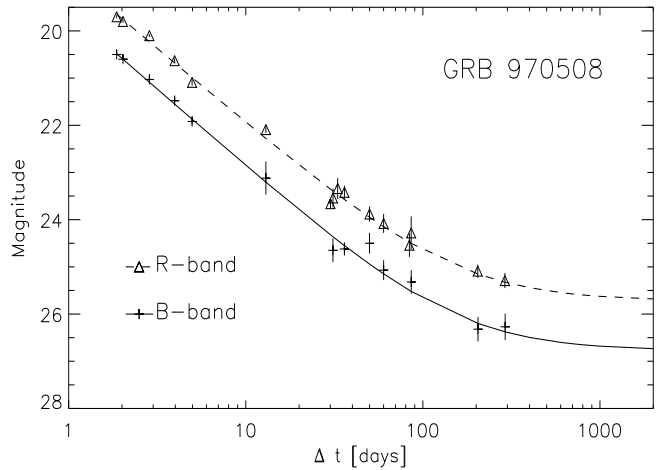,height=2.7in,width=3.8in}}
\caption[Light Curve of 970508]{Light curve of the optical transient
of GRB 970508.  Both $R$- (dashed; triangles) and $B$- band (solid;
crosses) data were compiled and transformed to a single photometric
system by Sokolov et al.~1998 (see references therein).  The latest
two data points on each light curve are from this paper.  The
constant flux of the underlying galaxy (the purported host) dominates
the light at late times.}
\label{fig:ltcurve}
\end{figure*}

To search for any potential offset of the OT and the galaxy, we used
an early image of the GRB field obtained on COSMIC at the Palomar
200-inch telescope on May 13.6 1997 UT while the transient was still
bright ($R$ $\approx$ 20 mag; see Djorgovski et al.~1997).  Assuming
the power-law behavior continued, the light at the transient position
is now dominated by the galaxy, with the transient contributing less
than 30\% to the total flux (see figure \ref{fig:ltcurve}).

We registered the Keck LRIS and the P200 COSMIC $R$-band (300-s)
images, by matching 33 relatively bright ($R \ale 23$ mag) objects in
a 4\arcmin $\times$ 4\arcmin\ field surrounding the GRB transient.
The coordinate transformation between the two images accounted for
pixel scale, rotating, translation, and higher-order distortion.  The
r.m.s.~of the transformed star positions (including both axes) was
$\sigma = 0.56$ LRIS pixels (= 0.121 arcsec).  We find the angular
separation of the OT and the galaxy to be $\ale 0.814$ pixels (= 0.175
arcsec), which includes the error of the transformation and centering
errors of the objects themselves.  The galaxy is found well-within 1.7
pixels = 0.37 arcsec (3-$\sigma$) of the OT.

\begin{figure*}[tbp]
\centerline{\psfig{file=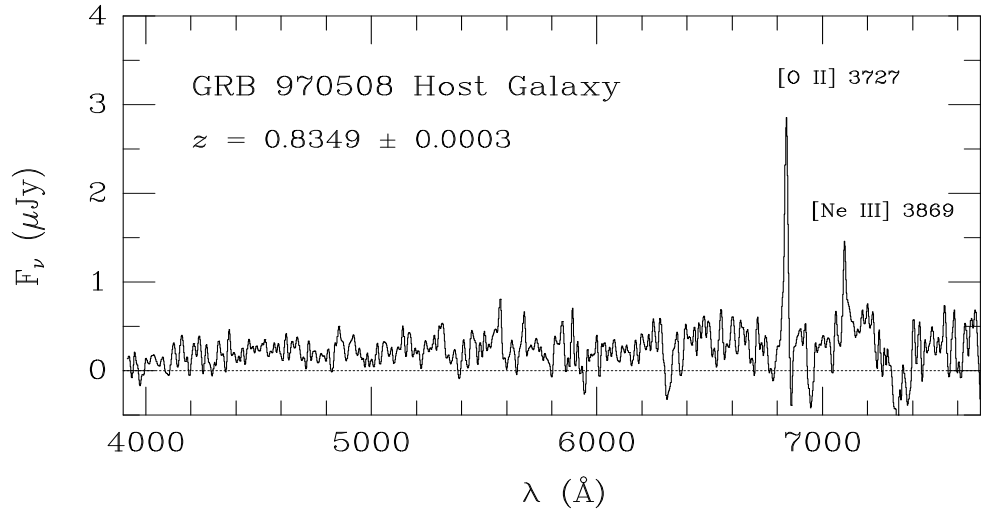,height=3.0in,width=3.2in}}
\caption[]{The weighted average spectrum of the host galaxy of GRB
970508, obtained at the Keck telescope.  The spectra were smoothed
with a Gaussian with a $\sigma = 5$\AA, roughly corresponding to the
instrumental resolution.  Prominent emission lines are labeled. }
\label{fig:thespect}
\end{figure*}

The averaged spectrum of the OT shows a very blue continuum, a
prominent emission line at $\lambda_{\rm obs} = 6839.7$ \AA, and a
somewhat weaker line at $\lambda_{\rm obs} = 7097.7$ \AA
(fig.~\ref{fig:thespect}).  We interpret the emission features as [O
II] 3727 \AA\ and [Ne III] 3869 \AA\ at the weighted mean redshift of
$z = 0.8349 \pm 0.0003$.  Our inferred redshift for the host is
consistent (within errors) with that of the absorbing system
discovered by Metzger et al.~(1997a).

The spectrum of the nearby galaxy g1 shows a relatively featureless,
blue continuum.  We are unable to determine its redshift at this
stage.
 
Our spectroscopic measurements give a magnitude $R \approx 25.05$ mag
(OT+host) at the mean epoch ($\approx 163$ days after the GRB) of our
observations, in excellent agreement with the magnitude inferred from
the fit to direct imaging data (see figure 1).

\section{Discussion}

After an initial brightening, the light-curve of the optical transient
did not deviate significantly from a power-law over first 100 days
after the burst (see also eg.~Galama et al.~1998; Garcia et
al.~1998). Assuming the blastwave producing the afterglow expanded
relativistically (bulk Lorentz factor, $\Gamma \age$ few) during the
beginning of the light-curve decline, the observed flux was produced
from within an angle $\omega_\Gamma \simeq 1/\Gamma$ of the emitting
surface.  As the blastwave expands, $\omega_\Gamma$ increases with
time. As long as the angle through which the blastwave is collimated
is greater than $\omega_\Gamma$, there would be no obvious break in
the light-curve (eg.~Sari, Piran, \& Narayan 1998).  One might expect,
in addition, the blastwave to eventually become sub-relativistic
resulting not only in a larger observed surface area, but perhaps a
change in surface emmisivity. Curiously, an apparent break expected in
either scenario did not materialize.

The spatial coincidence of the transient and the underlying galaxy may
simply be a chance projection of the transient, which lies beyond
$z=0.835$, and the galaxy at $z=0.835$.  The surface density of
galaxies down to $R=25.7$ mag is 48.3 per arcmin$^{2}$. It is
important to note that we know {\it a priori} that the host must lie
in the redshift range $0.835 \ale z \ale 2.1$ (Metzger et al.~1997).
The fraction of galaxies within this range is $\sim 50$ percent of the
total at the magnitude level (Roche et al.~1996).  The {\it a
posteriori} Poisson probability of finding such a galaxy with in 0.37
arcsec from the OT is 3 $\times$ 10$^{-3}$.  Keeping in mind the
limitations of {\it a posteriori} statistics, this small probability
and the trend that GRB transients appear to be nearly spatially
coincident with galaxies lead us to suggest this galaxy is the host of
GRB 970508.

Assuming a standard Friedman model cosmology with $H_0 = 65$ km
s$^{-1}$ Mpc$^{-1}$ and $\Omega_0 = 0.2$ we derive a luminosity
distance of $1.60 \times 10^{28}$ cm to the host galaxy. The observed
equivalent width in the [O II] line is $(115 \pm 5)$ \AA, or about 63
\AA\ in the galaxy's restframe.  However, this also includes the
continuum light from the OT at this epoch.  Correcting for the OT
contribution would then double these values of the equivalent width.
This is at the high end of the distribution for the typical field
galaxies at comparable magnitudes and redshifts (Hogg et al.~1998).

The implied [O II] line luminosity, corrected for the extinction, is
$L_{3727} = (9.6 \pm 0.7) \times 10^{40}$ erg s$^{-1}$.  Using the
relation from Kennicutt (1998) we estimate the star formation rate
(SFR) $\approx 1.4 ~M_\odot ~{\rm yr}^{-1}$.

An alternative estimate of the SFR can be obtained from the continuum
luminosity at $\lambda_{rest} = 2800$ \AA\ (Madau, Pozzetti, \&
Dickinson~1998).  The observed, interpolated continuum flux from the
host itself (i.e., not including the OT light) at the corresponding
$\lambda_{obs} \approx 5130$\AA, is $F_\nu \approx 0.11 ~\mu$Jy,
corrected for the estimated Galactic extinction ($A_V \approx 0.08^m$;
Djorgovski et al.~1997).  For our assumed cosmology, the restframe
continuum luminosity is then $L_{2800} \approx 1.93 \times 10^{27}$
erg s$^{-1}$ Hz$^{-1}$, corresponding to SFR $\approx 0.25 ~M_\odot
~{\rm yr}^{-1}$.  This is notably lower than the SFR inferred from the
[O II] line.  We note, however, that neither is known to be a very
reliable SFR indicator.  Both are also subject to the unknown
extinction corrections from the galaxy's own ISM (the continuum
estimate being more sensitive).  We thus conclude that the {\it lower
limit} to the SFR in this galaxy is probably about 0.5 to 1 $M_\odot
~{\rm yr}^{-1}$.

The observed flux in the [Ne III] 3869 \AA\ line is $F_{3869} = (1.25
\pm 0.1) \times 10^{-17}$ erg cm$^{-2}$ s$^{-1}$, not corrected for
the extinction.  The flux ratio of the two emission lines is
$F_{3869}/F_{3727} = 0.44 \pm 0.05$.  This ratio is about 10 times
higher than the typical values for H II regions.  Nonetheless, it is
in the range of photoionization models for H II regions by Stasinska
(1990), for different combinations of model parameters, but generally
for effective temperatures $T_{eff} \geq 40,000$ K.

The inferred host luminosity is in agreement with the upper limit from
earlier {\it HST} observations (Pian et al.~1998).  Further, our
derived $B$- and $R$-magnitudes for the galaxy correspond to a
continuum with a power law $F_\nu \sim \nu^{-1.56}$.  Extrapolating
from the observed $R$-band flux to the wavelength corresponding to the
restframe $B$-band (about 8060 \AA), and correcting for the Galactic
foreground extinction, we derive the observed flux $F_\nu(\lambda =
8060 \AA) \approx 0.22 \mu$Jy.  For our assumed cosmology, the implied
restframe absolute magnitude is then $M_B \approx -18.55$.  Thus the
restframe $B$-band luminosity of the host galaxy is about $0.12 L_*$
today.

This galaxy is roughly 2 magnitudes fainter than the knee of the
observed luminosity function of all galaxies between redshift $z=0.77$
and 1.0 (Canada-France redshift survey; Lilly et al.~1995) and one
magnitude fainter than late-time, star-forming galaxies in the 2dF
survey (Colless 1998).  The specific SFR per unit luminosity is
high. This object can thus be characterized as an actively
star-forming dwarf galaxy.  Objects of this type are fairly common at
comparable redshifts.

\section{Conclusions}

The high effective temperature implied by the relative line strengths
of [Ne III] and [O II], suggests the presence of a substantial
population of massive stars and thus, active and recent star
formation.  This, in turn, gives additional support to the ideas that
the origin of GRBs is related to massive stars (e.g.~Wijers et
al.~1998; Totani 1998).

An alternative possibility for the origin of the [Ne III] 3869 \AA\
line is photoionization by a low-luminosity AGN.  While we cannot
exclude this possibility, we note that there is no other evidence in
favor of this hypothesis, and moreover we see no other emission lines,
e.g., Mg II 2799 \AA, which would be expected with comparable
strengths in an AGN-powered object.

What may be surprising, in the neutron star binary (NS--NS) model of
GRB progenitors (eg.~Narayan, Paczy\'nski, \& Piran 1992; Paczy\'nski
1986), is that GRB 970508 appears so close ($< 2.7$ kpc) to a dwarf
galaxy ($L \approx 0.1 L_*$).  Bloom, Sigurdsson, \& Pols (1998)
recently found that less than 15 percent of NS--NS binaries will merge
within 3 kpc of a comparable under-massive galaxy.  If GRBs are
consistently found very near ($\ale$ few kpc) of their purported host,
then progenitor models such as microquasars (Paczy\'nski 1998),
``failed'' Type Ib supernovae (Woosley 1993), or black hole--neutron
star binaries (Mochkovitch et al.~1993; M\'esz\'aros \& Rees 1997),
all of which are expected to produce GRBs more tightly bound to their
hosts, would be favored.

\acknowledgements

It is a pleasure to thank S.~Odewahn, M.~van Kerkwijk, R.~Gal, and
A.~Ramaprakrash for assistance during observing runs at Keck, P.~Groot
for comments, and R.~Sari for helpful discussions concerning
inferences from the light-curve.  SRK's research is supported by the
National Science Foundation and NASA.  SGD acknowledges a partial
support from the Bressler Foundation.

\newpage

\begin{table*}[tb]
\caption{Late-Time GRB 970508 Imaging and Spectroscopic
Observations}
\label{tab:obs}
\begin{tabular}{lcccccccr}
\hline\hline
\multicolumn{1}{c}{Date}& Band/ & 
\multicolumn{1}{c}{Int.~Time} &  \multicolumn{1}{c}{Seeing} & $\Delta t$ &
\multicolumn{2}{c}{Object Magnitude} & Observers\\

 UT & Grating & (sec) & (arcsec) & (days) &
\multicolumn{1}{c}{observed$^a$} & \multicolumn{1}{c}{OT
predicted$^b$}\\ \hline

Nov.~28, 1997 & R & 5400 & 1.2 & 203.8 & 25.09$\pm 0.14^c$ & 25.63$\pm 0.2$
            & Kulkarni, van Kerkwijk, \& \\

Nov.~29, 1997 & R & 600 & 1.2 & 204.8  & & & Bloom \\
             
Nov.~29, 1997 & R & 600 & 1.1 & 204.8  & & & `` \\
             
Nov.~29, 1997 & B & 2400 & 1.1 & 204.8 & 26.32$\pm 0.26$ & 
26.65$\pm 0.25$ & `` \\

Feb.~22, 1998 & R & 2400 & 1.1 & 290.5 & 25.29$\pm 0.16$ & 
26.07$\pm 0.2$ & Kulkarni, Ramaprakash, \& \\
             
Feb.~23, 1998 & B & 2400 & 1.2 & 291.5 & 26.27$\pm 0.28$ & 
27.11$\pm 0.25$ & van Kerkwijk \\

\hline

Oct.~3, 1997 & 300 & 5400 & 0.8 & 147.8 & & & Djorgovski, Odewahn \\
Nov.~2, 1997 & 300 & 5400 & $< 1.0$ & 177.8   &  & & Djorgovski, Gal   \\

\hline
\end{tabular}
\raggedright

{\footnotesize 

\noindent $^a$ Magnitudes are derived from V.~Sokolov's tertiary
reference stars.  The (conservative) 1-$\sigma$ errors include
statistical uncertainties in the reference transformation and the
OT+host itself.  All nights (imaging observations) were photometric.

\noindent $^b$ Predicted magnitudes are derived from a pure power-law
decline using light curve data from $\Delta t \ale 100$ days (Sokolov
et al.~1998).  Errors are estimated using the uncertainties in both
the magnitude scaling and power-law index.

\noindent $^c$ The $R$-band magnitude quoted is derived from the sum
of R-band images over two nights.

}
\end{table*}

\end{document}